\documentclass{emulateapj}
\usepackage{amsmath}
\usepackage{graphicx}
\usepackage{amsfonts}
\usepackage{natbib}
\usepackage{color}
\usepackage{epstopdf}
\usepackage{graphicx}
\usepackage{multirow}

\begin{document}

\title[The X-ray QPO in 1H 0707-495]{Two transient X-ray Quasi-Periodic Oscillations separated by an intermediate state in 1H 0707-495}
\author{Peng-fei Zhang\altaffilmark{1,2}, Peng Zhang\altaffilmark{1,3}, Neng-hui Liao\altaffilmark{1}, Jing-zhi Yan\altaffilmark{1}, Yi-zhong Fan\altaffilmark{1,4},  and Qing-zhong Liu\altaffilmark{1,4}}
\altaffiltext{1}{Key Laboratory of Dark Matter and Space Astronomy, Purple Mountain Observatory, Chinese Academy of Sciences, Nanjing 210008, China;
zhangpengfee@pmo.ac.cn, jzyan@pmo.ac.cn, yzfan@pmo.ac.cn, qzliu@pmo.ac.cn}
\altaffiltext{2}{Key Laboratory of Astroparticle Physics of Yunnan Province, Yunnan University, Kunming 650091, China}
\altaffiltext{3}{University of Chinese Academy of Sciences, Beijing, 100012, China}
\altaffiltext{4}{School of Astronomy and Space Science, University of Science and Technology of China, Hefei, Anhui 230026, China}

\begin{abstract}
In the narrow-line Seyfert 1 galaxy 1H 0707-495, recently a transient quasi-periodic oscillation (QPO) signal with a frequency of $\sim 2.6\times 10^{-4}$ Hz
has been detected at a high statistical significance.
Here, we reanalyze the same set of XMM-Newton data observed on 2008 February 4 with the Weighted-Wavelet Z-transform (WWZ) method.
In addition to confirming the previous findings,
we also find another QPO signal with a frequency of $\sim 1.2\times 10^{-4}$ Hz in a separated X-ray emission phase at the significance level
of $\sim 3.7\sigma$. The signal is also found fitting an auto-regressive model though at a lower significance.
The frequency ratio between these two signals is $\sim 2:1$. The analysis of other XMM-Newton measurements of 1H 0707-495 also
reveals the presence of the $\sim 2.6\times 10^{-4}$ Hz ($\sim 1.2\times 10^{-4}$ Hz) QPO signal on 2007 May 14 (2010 September 17)
at the significance level of $\sim 4.2\sigma$ ($\sim 3.5\sigma$).
The QPO frequency found in this work follows the $f_{QPO}-M_{BH}$ relation reported in previous works spanning from stellar-mass to supermassive black holes.
This is the first time to observe two separated transient X-ray QPO signals in active galactic nuclei (AGNs), which sheds new light on the physics of accreting supermassive black holes.
\end{abstract}

\bigskip
\keywords{ galaxies: active - galaxies: nuclei - galaxies: individual (1H 0707-495) - X-rays: galaxies}
\bigskip
\section{INTRODUCTION}
\label{sec:intro}
Active galactic nuclei (AGNs), the most persistent luminous sources of electromagnetic radiation in the universe,
are widely believed to be powered by the accretion of material onto supermassive black holes (SMBHs).
Narrow-line Seyfert 1 galaxies (NLS1s) serve as a peculiar subclass of AGNs,
with common definition based on their optical spectrum (a narrow width of the broad Balmer emission lines with FWHM (H$\beta$) $<$ 2000 km $\rm s^{-1}$,
along with strong optical $\rm Fe_{II}$ lines and weak forbidden lines) \citep{Osterbrock1985,Goodrich1989,Pogge2000}.
Other extreme properties include rapid X-ray variability, near-Eddington accretion rates as well as strong star formation of the host galaxies,
indicating a rapidly growing phase of their central black holes  \citep{Komossa2008}.
Furthermore, the $\gamma$-ray emissions of NLS1s \citep{Abdo2009,Liao2015} render them a new member of the jetted-AGN family.
Therefore, the studies of NLS1s are essential to have a through understanding of the AGN phenomena.

Periodic emissions are well detected in X-ray light curves of Galactic binary systems
\citep{Bolton1972,Webster1972,Ackermann2012} and
the periodic variability studies can be powerful diagnostic tools \citep{Remillard2006}.
However, in AGNs, the periodicity is relatively rare.
Some detections or evidences for periodic variabilities  in optical, X-ray, and/or gamma-ray emission of AGN have been reported
in the literature \citep{Kidger1992,Valtonen2006,1553,Reis2012,Zhang2017a,Covino2017}.
In X-rays, a significant transient QPO has been detected in NLS1 galaxy RE J1034+396 \citep{Gierlinski2008}.
Recently, the other significant transient QPO signal has been reported in NLS1s galaxy 1H 0707-495 \citep{Pan2016} and Mrk 766 \citep{Zhang2017b}.
In this work, we re-analyze the data set adopted by \citet{Pan2016} with the Weighted-Wavelet Z-transform method.
In addition to confirm previous finding, we find a new QPO signal in the early part of the light-curve with a period cycle
of $\sim$ 8240 s at the confidence level of $\sim 3.7\sigma$.
These two QPO signals, with a frequency ratio of $\sim2:1$, are separated by an intermediate state.
Interestingly, in two other observation data sets these signals appeared, though individually and at lower significance levels.
Our conclusions are thus further strengthened.

\section{Observation and Data Analysis}
\label{sec:Observations}

\subsection{The construction of X-ray light curves}
\label{subsec:make lc}
The X-ray Multi-Mirror Mission (XMM-Newton) was launched on December 10th 1999 by European Space Agency's (ESA).
 {\it XMM-Newton} carries a set of three X-ray CCD cameras (EPIC), including two MOS \citep{Turner2001} and one PN \citep{Struder2001},
with an unprecedented large effective area.
The NLS1 galaxy 1H 0707-495 had been monitored 15 times over 40 ks with the former three detectors between 2000 January and 2011 February in a full frame imaging mode.
The data analysis is performed following the standard procedure in the Science Analysis Software (SAS) with version of 16.0.0 provided by the {\it XMM-Newton} science operations center \footnote{https://www.cosmos.esa.int/web/xmm-newton/sas-threads}.
The events are extracted from a circle region of interest (ROI) of 40-arcsec radius
centered on the position (R.A. = 107.173, dec. = -49.552) of the target in 0.2-10.0 keV with time-bin of 100 s.
We exclude the events in periods with high background flaring rates exceeding 0.4 counts/s,
and only the good events (the PATTERN $\le$ 4 for PN and PATTERN $\le$ 12 for MOS) are used in generating light curves.
For background, the events are extracted from a source-free circle (without any other source) with the same diameter in the same chip.
Then, the correction performed by tool {\it epiclccorr} has been accomplished, and the pile-up effect is negligible.
We combine the three {detectors (PN+MOS1+MOS2)} background subtracted light curves (we denote it as EPIC light-curve),
and show them in the top panels of Fig. \ref{wwz} and Figs. \ref{20100917}-\ref{20070514}. Apart from that,
there are two identical Reflection Grating Spectrometers (2RGS) \citep{Herder2001} onboard on {\it XMM-Newton}.
To confirm the former results, we also derive the light curves from 2RGS products with tool {\it rgslccorr} following the standard procedure and
show the results in Figs. \ref{wwz_rgs}-\ref{20070514}.
And the following analysis is based on these light curves.

\subsection{The analysis of periodic variability}
\label{subsec:Search qpo}
\begin{figure*}
\centering
\includegraphics[scale=0.32]{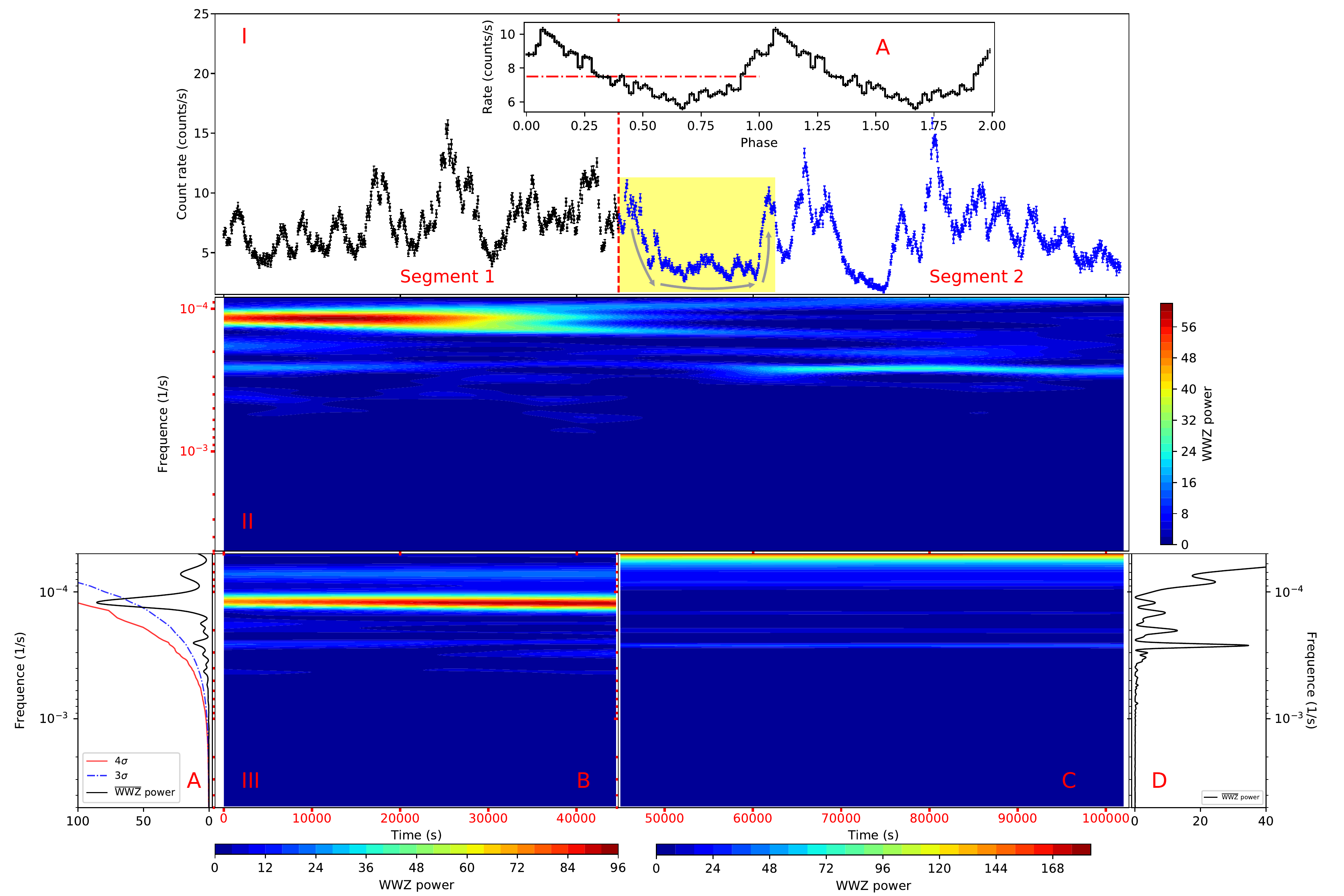}
\caption{Panel I: The combined light-curve from PN, MOS1 and MOS2 detectors in 0.2-10 keV with 100-s per bin is observed on 2008 February 4 (Obs ID:0511580401),
                  the two segments are separated with a red dashed line.
                  And the pulse shape (the folded X-ray light-curve for segment 1) with a period cycle of 8244.36 s is shown
                  in the inset A panel (two cycles are shown). The yellow box represents the intermediate state.
		 Panel II: The 2D plane contour plot of the WWZ power of the whole X-ray light-curve with color contour filled.
		               The color-bar is scaled with WWZ power value.
		 Panel III: in the A part, The black solid line represents time-averaged WWZ power, the curves of confidence levels of $4\sigma$ and $3\sigma$
		                are shown with red solid and blue dotted-dashed lines, respectively,
		                basing on the null hypothesis with $\alpha \sim 2.4$ and $Log(f_{bend}) \sim -3.69$;
		                The B part presents the contour plot of the WWZ power of segment 1;
                                 The C part is the contour plot of the WWZ power of segment 2;
		                The D part gives the time-averaged WWZ power of segment 2.}
\label{wwz}
\end{figure*}

The wavelet transform is a widely used method for searching the periodic signal in the time series
by changing the parameters of wavelets to fit light curves in both time and frequency domains.
This method is different from Fourier analysis, which pays attention to a limited time span of data.
A key advantage it has over Fourier transforms is temporal resolution: it captures both frequency and location information (location in time).
However, this method may yield untrustworthy results because of the local number density of the uneven data points.
A modified version of weighted wavelet Z-transform with a Morlet mother function has been provided in \citet{Foster1996},
which handles efficiently this problem by rescaling the wavelet function. Therefore, we calculate WWZ power spectral for light curves.
In the II and III panels of Fig.\ref{wwz}, we present the color-scaled WWZ power spectrum for the whole light-curve;
WWZ power spectral (WWZ power develops and evolves in frequency and amplitude over time, the detailed information provided by \citealt{Foster1996}.)
and time-averaged WWZ power spectral (the average WWZ power at a given frequency along time) for the first and second segment, respectively.

\citet{Pan2016} divided the light-curve into two segments (see Fig. 1 therein) and then calculated the power spectrum density (PSD) with Fourier analysis
for the second segment. They detected a QPO with a strong peak at $(2.6\pm 0.18) \times10^{-4}$ Hz (which corresponds to a period of 3800 s).
 We use the WWZ method to re-analyze the data and surprisingly find two signals, separated by a short intermediate state, at different frequencies
(see the panel II of Fig.\ref{wwz}).
We thus calculate the power spectrum for both segments, respectively.
The WWZ powers for the second segment reveal a clear peak (see the bottom C and D panels of Fig.\ref{wwz}), in agreement with \citet{Pan2016}.
 While, in the segment 1, another QPO also appears in the WWZ power and time-averaged WWZ at $(1.21\pm0.20)\times10^{-4}$ Hz
(see the III A and III B panels of Fig.\ref{wwz}).

The uncertainty is evaluated as the means of the full width at half maximum (FWHM) fit Gaussian function
centered around the maximum time-averaged WWZ power.
The maximum time-averaged WWZ power is $\sim14.5$ times of the underlying continuum with the false-alarm probability \citep[FAP;][]{Vaughan2005} of $\sim2.5\times10^{-7}$
for the combined {EPIC} lightcurve, and for the combined 2RGS light-curve, it is 10.5 times than the underlying continuum with the FAP of $1.4\times10^{-5}$.
In order to establish a robust significance, we generate the artificial
light curves with a simulator \citep{Timmer1995,Emmanoulopoulos2013} for all light-curves reported in this paper, and the simulator is based on
the power spectral density (PSD) and the probability density function of observed variation.
Therefor, the artificial light curves have full properties of statistical and variability as the X-ray flux of the target.
For determining the PSD best-fit, we fit the PSD with a bending power law plus constant function (null hypothesis) as $P(f)~=~Af^{-1}(1+{(f/f_{bend})}^{\alpha-1})^{-1}+C$
employing maximum likelihood method.
In the likelihood method, we use likelihood function of $\mathcal{L}=\prod_{j=1}^{N-1}p(I_j|S_j)=\prod_{j=1}^{N-1}\frac{1}{S_j}\exp(-I_j/S_j)$,
where $I_j=I(f_j)$ representing the power at frequency $f_j=j/(N\times\Delta T)$ and $S_j$ representing the expectation value at $f_j$ basing on the null hypothesis
\citep[see][for the details]{Groth1975,Leahy1983,Barret2012}.
For the model, the free parameters $\alpha$, $f_{bend}$, $C$ and $A$ are the spectral index above the bend,
the bend frequency, the constant (Poisson noise) and the normalization, respectively \citep{Markowitz2003,McHardy2006,Gonzales2012,Kelly2014}. 
The detailed information of the method for evaluating of significance level is provided
in \cite{Gierlinski2008,Emmanoulopoulos2013,1553,Bhatta2016}. In total, we generate $\sim~2\times10^{5}$ artificial light curves (the same number for others).
In the III A panel of Fig.\ref{wwz}, the $4\sigma$ and $3\sigma$ confidence levels are shown with the red solid and blue dotted-dashed curves, respectively.
The significance curves represent the distribution of heights of the most significant peaks across any of the sampled frequencies basing the null hypothesis.
We compute the probability of obtaining the power peaks using {EPIC} and 2RGS light-curves.
The probabilities for the signal are $\sim99.98\%$ ($3.7\sigma$) and $\sim99.96\%$ ($3.5\sigma$), respectively.
{On 2010 September 17, the same signal is also detected, which is shown in the Fig. \ref{20100917}. The results of EPIC and 2RGS light-curves are shown in left and right images, respectively.
For EPIC the confidence level of the signal at $\sim (1.21\pm0.12)\times10^{-4}$ Hz is $\sim 3.5\sigma$ (99.95\%), and for 2RGS it is $3.2\sigma$ (99.87\%).
The probability of chance fluctuation in the EPIC power spectra from two independent observations
light-curves at $1.21\times10^{-4}$ Hz is $< 8.6\times10^{-8}$ (the corresponding significance is $\sim5.3\sigma$).}
Considering 15 times monitoring over 40 ks, the total exposure time is $\sim$ 1.3 Ms
meaning $\sim20$ segments of similar length presented here.
Including the results reported above, we detect QPO signals at $\sim (1.21\pm0.12)\times10^{-4}$ Hz in two segments.
While the number of trials of frequency bins is $5$ within FWHM of the power peak
(from $\sim1.1\times10^{-4}$ to $\sim1.3\times10^{-4}$ Hz, in fact the maximum of power spectra is independent of frequency bins).
Accounting the total number of trials {of $190\times5$}, the combined confidence level of the signal is $\sim4.0\sigma$ (99.993\%).

\begin{figure*}
\centering
\includegraphics[scale=0.5]{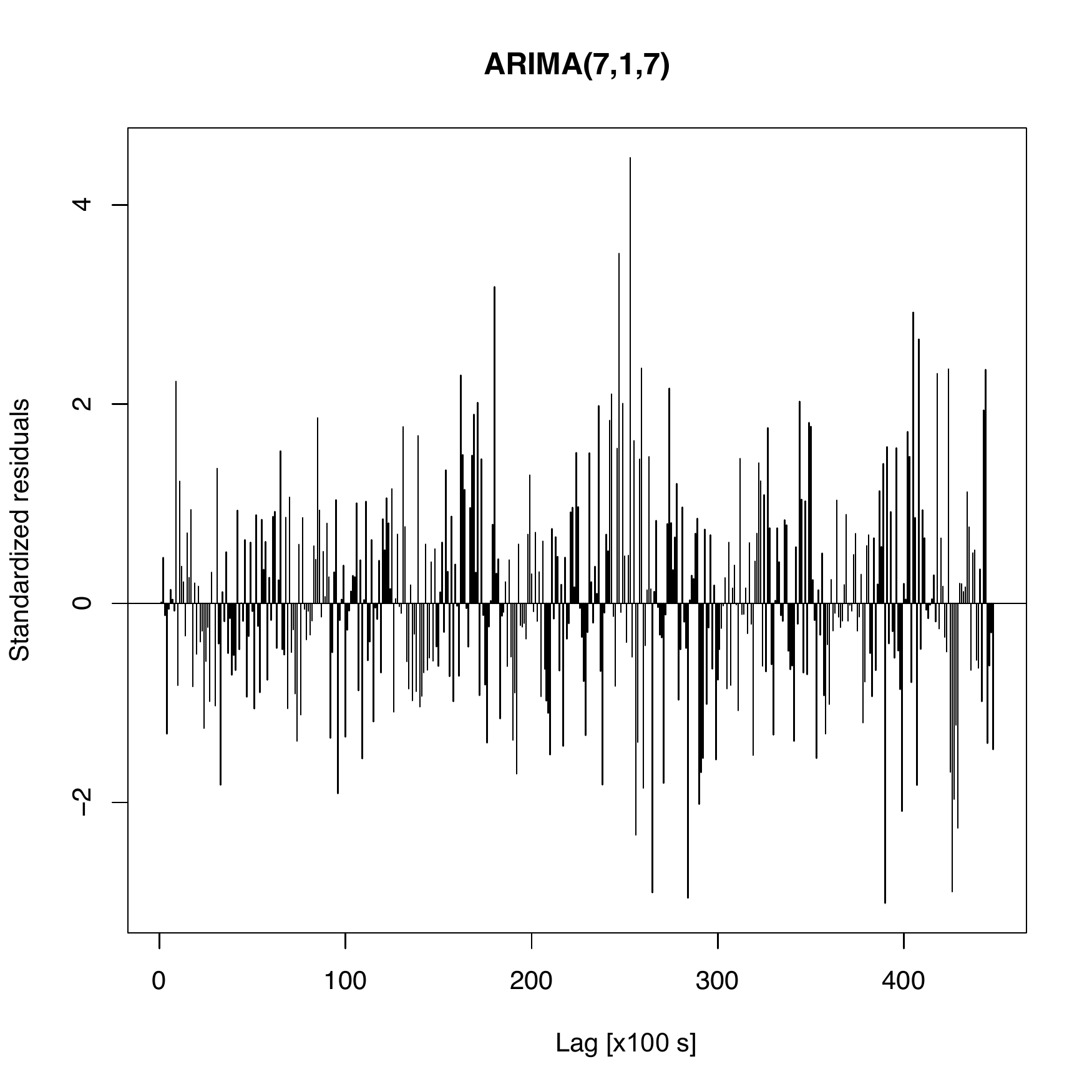}
\includegraphics[scale=0.5]{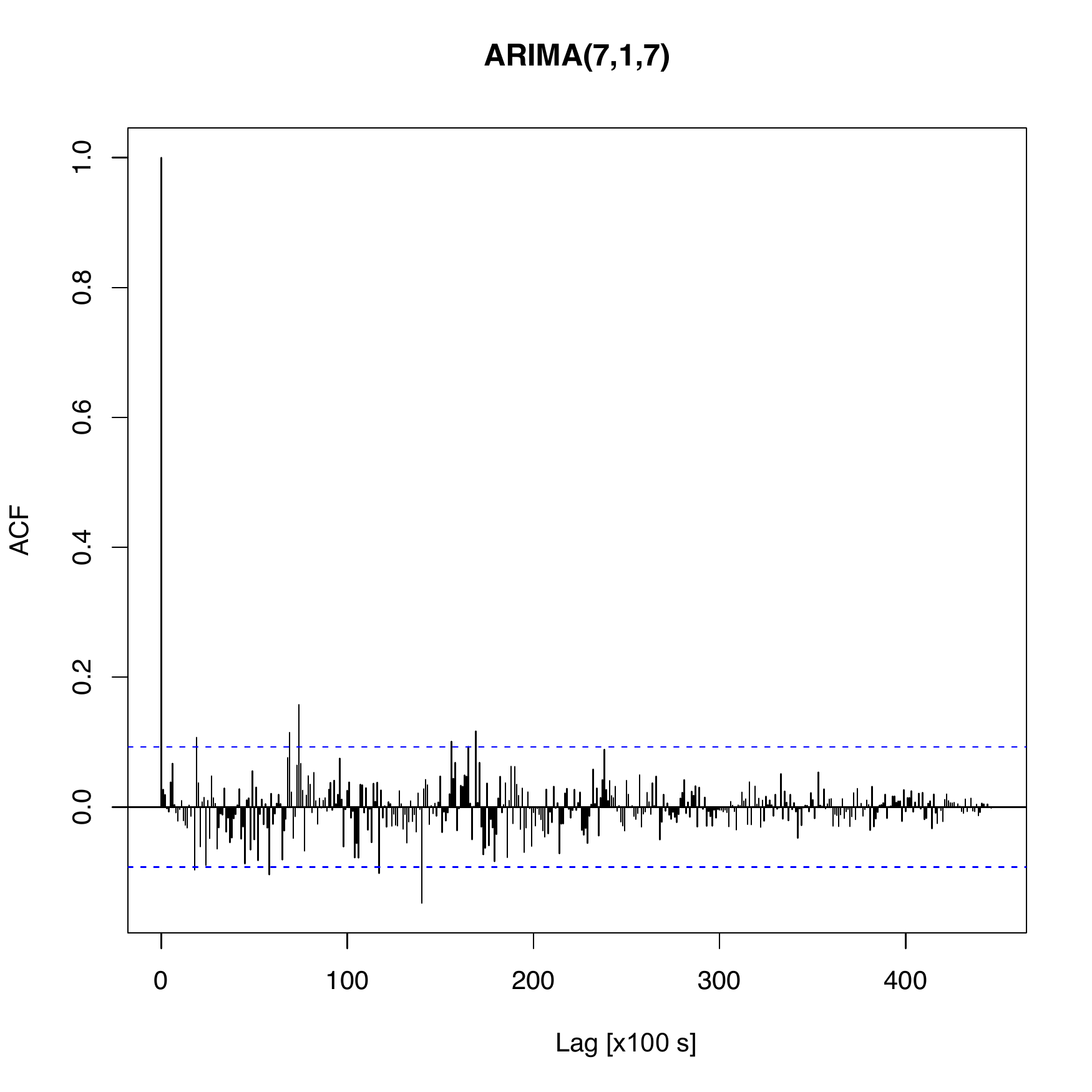}
\caption{Results of the ARIMA model fitting for segment 1 light curve. Left panel: the standard residuals of ARIMA(7, 1, 7) model fitting.
              Right panel: the residuals ACF of ARIMA(7, 1, 7) model fitting, where the blue dashed line represents the 2$\sigma$ confidence level.}
\label{arima}
\end{figure*}

\begin{figure*}
\centering
\includegraphics[scale=0.6]{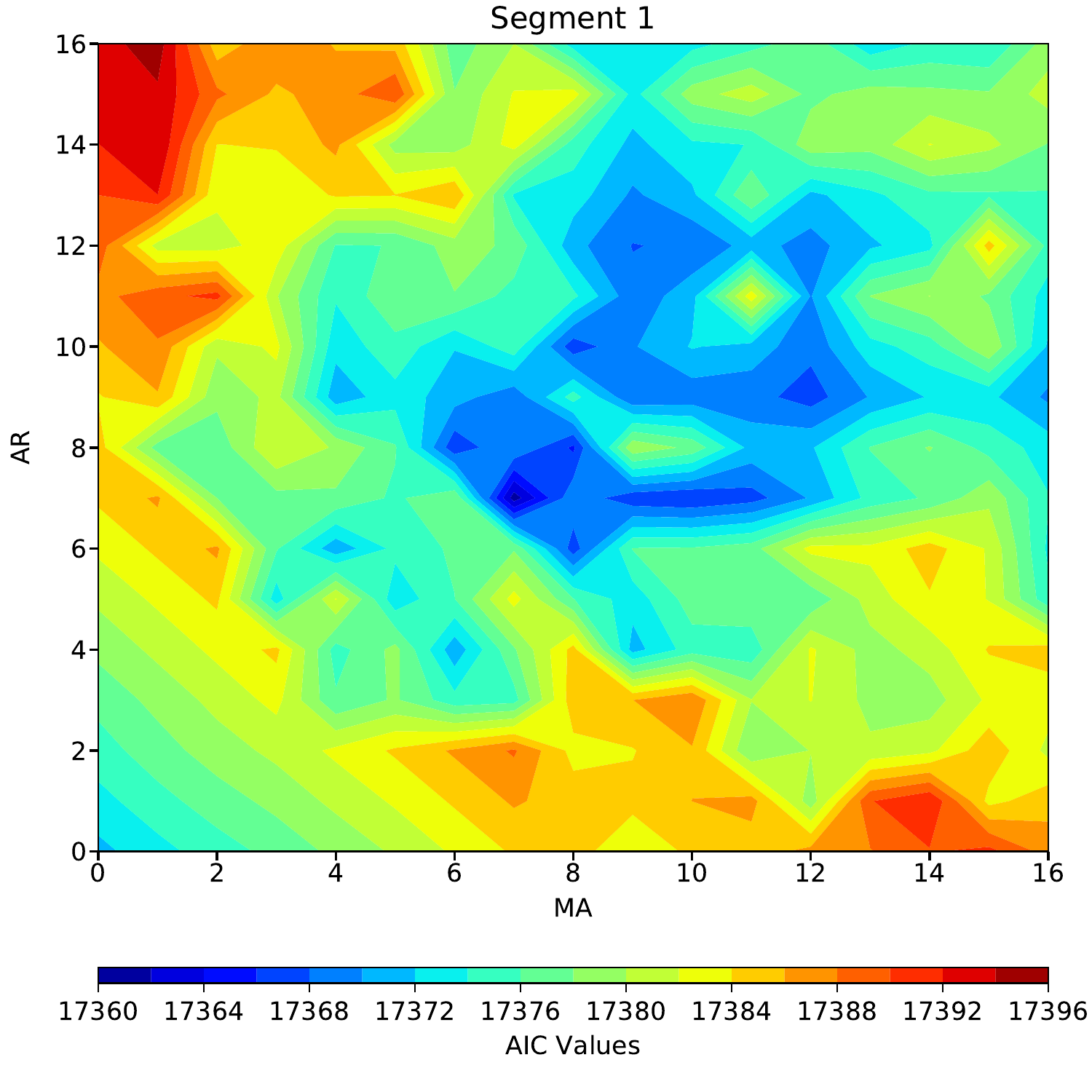}
\caption{The AIC for the ARIMA models fitting the segment 1 light curve. The color represents the AIC values.}
\label{arima_aic}
\end{figure*}

\begin{figure*}
\centering
\includegraphics[scale=0.4]{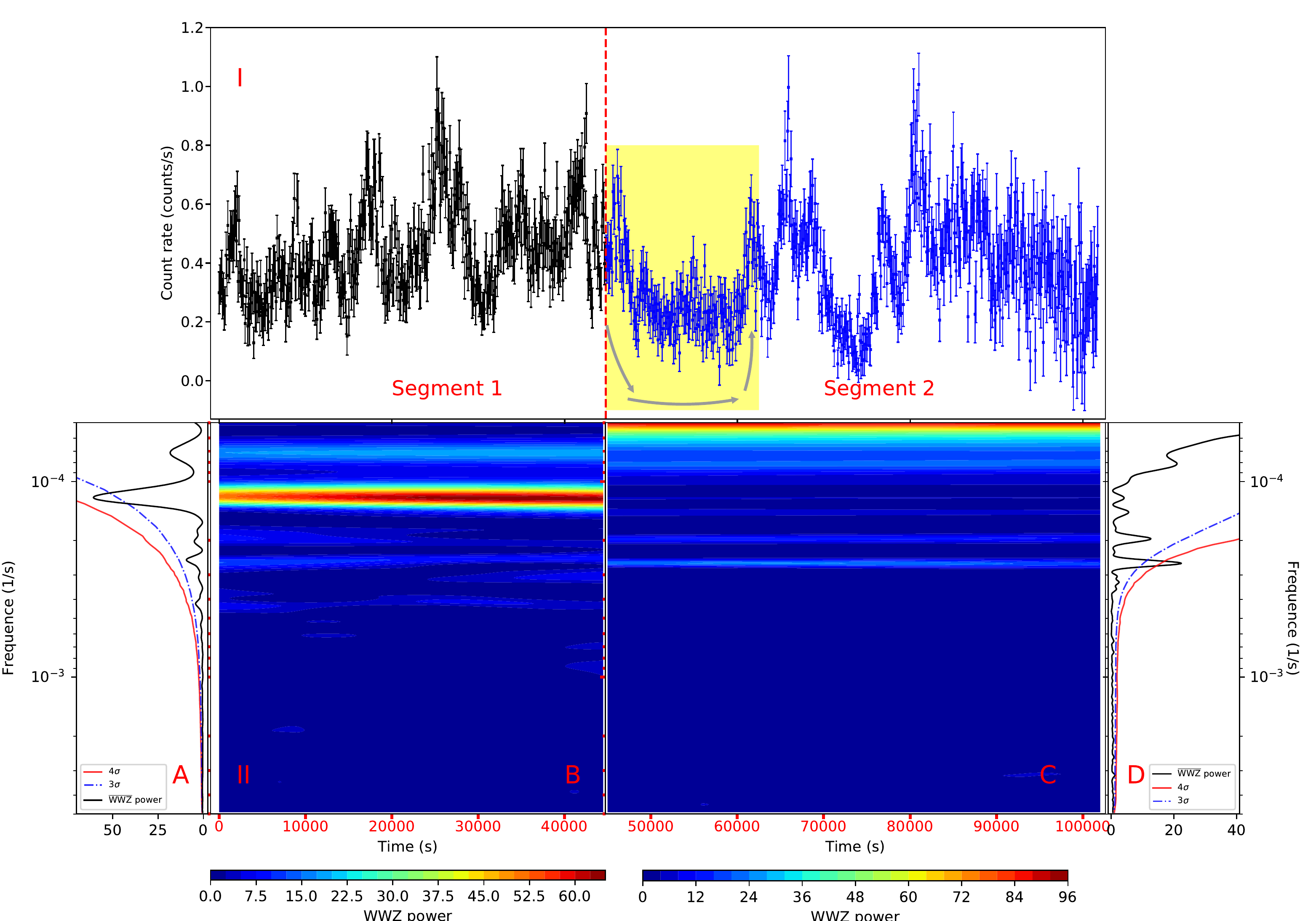}
\caption{In the upper panel, the 2RGS light-curve in 0.2-10 keV with 100-s per bin is observed on 2008 February 4.
              The parameters $\alpha$ and $Log(f_{bend})$ of the bending power-law are  $\sim 2.0$ and $\sim -3.69$ for segment 1
              and $\sim 5.0$ and $\sim -3.69$ for segment 2. Other instructions are same as Fig. \ref{wwz}.}
\label{wwz_rgs}
\end{figure*}

\begin{figure*}
\centering
\includegraphics[scale=0.35]{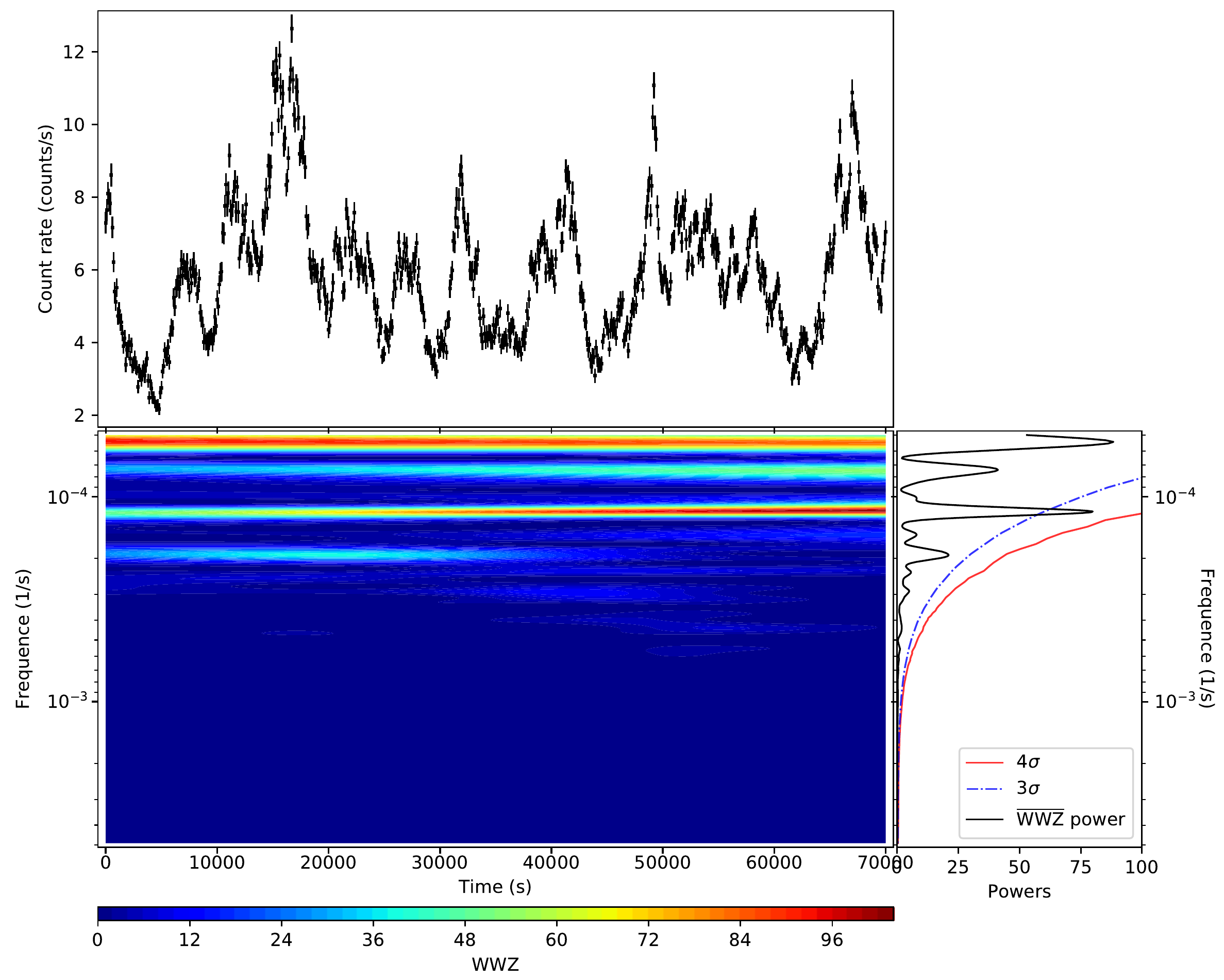}
\includegraphics[scale=0.35]{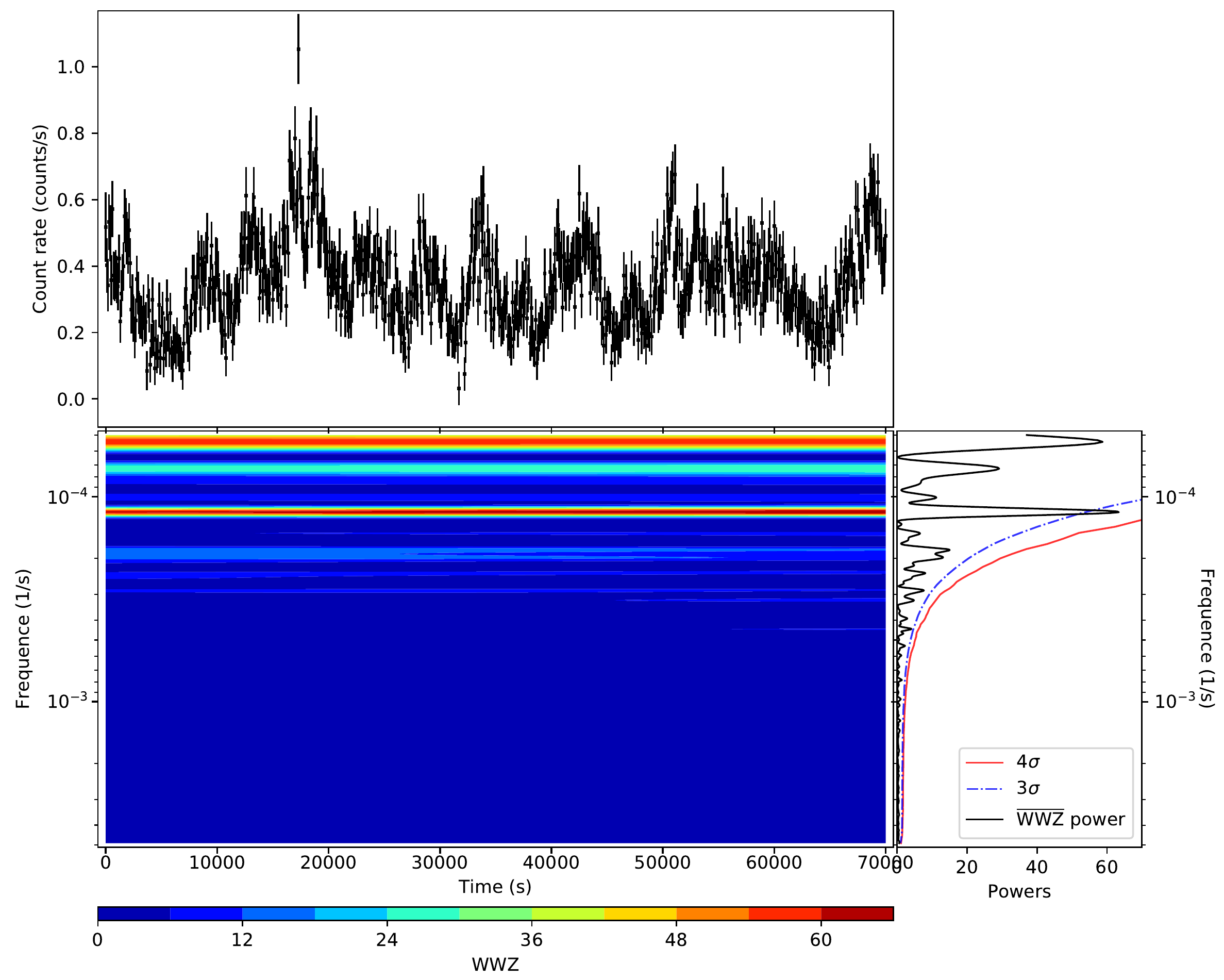}
\caption{For the both images, in the upper panel, the light-curves in 0.2-10 keV with 100-s per bin are
              the observations of XMM-Newton on 2010 September 17 (Obs ID: 0653510501), left image for EPIC and right image for 2RGS.
              The lower left panel is the contour plot of the WWZ power of upper light-curve.
              The black solid line represents time-averaged WWZ power, The confidence curves of $4\sigma$ and $3\sigma$
              are shown with red solid and blue dotted-dashed lines, respectively.
              The parameters $\alpha$ and $Log(f_{bend})$ of the bending power-law are  $\sim 2.4$ and $\sim -3.9$ for EPIC light-curve
              and $\sim 2.5$ and $\sim -3.9$ for 2RGS light-curve.}
\label{20100917}
\end{figure*}

\begin{figure*}
\centering
\includegraphics[scale=0.35]{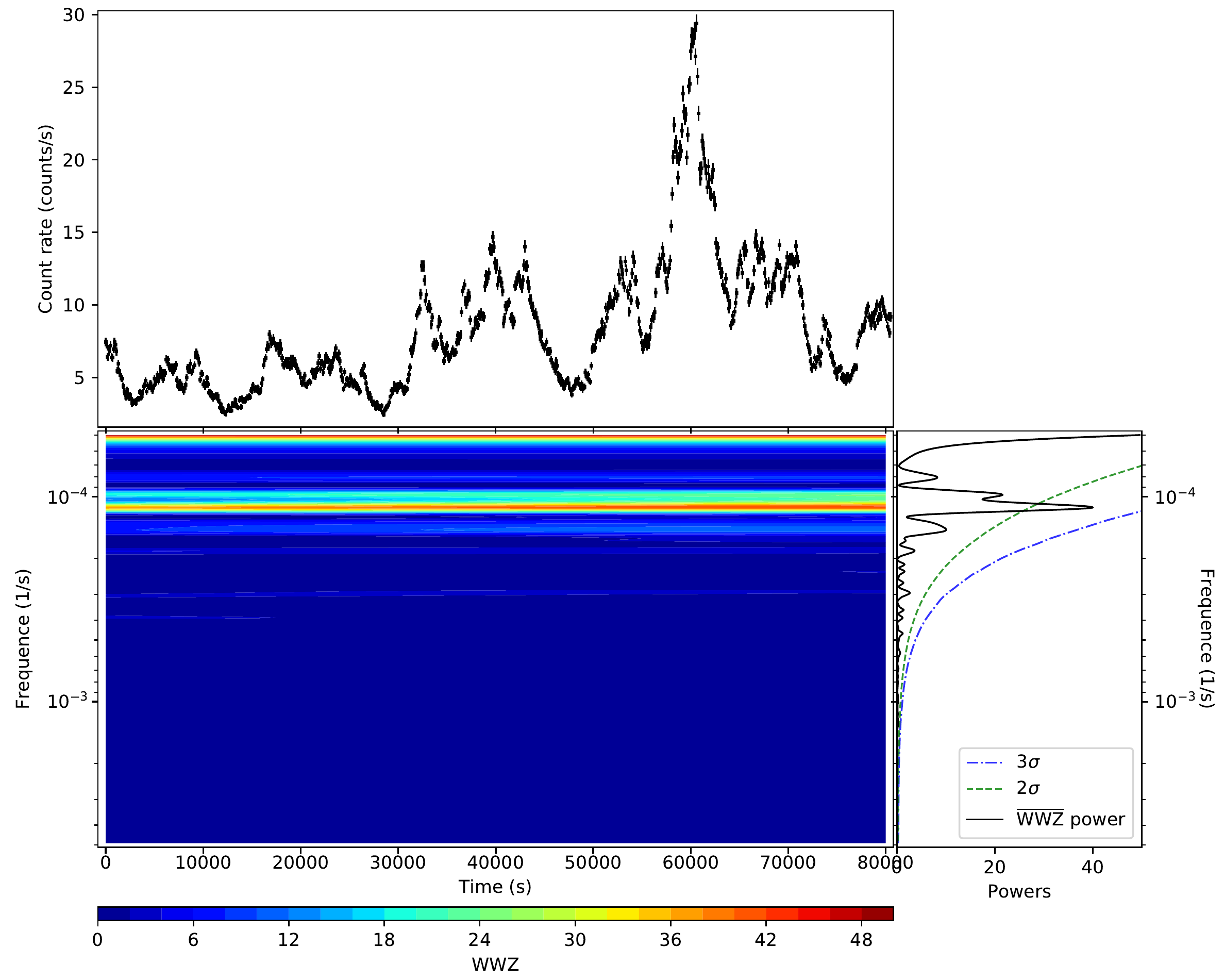}
\includegraphics[scale=0.35]{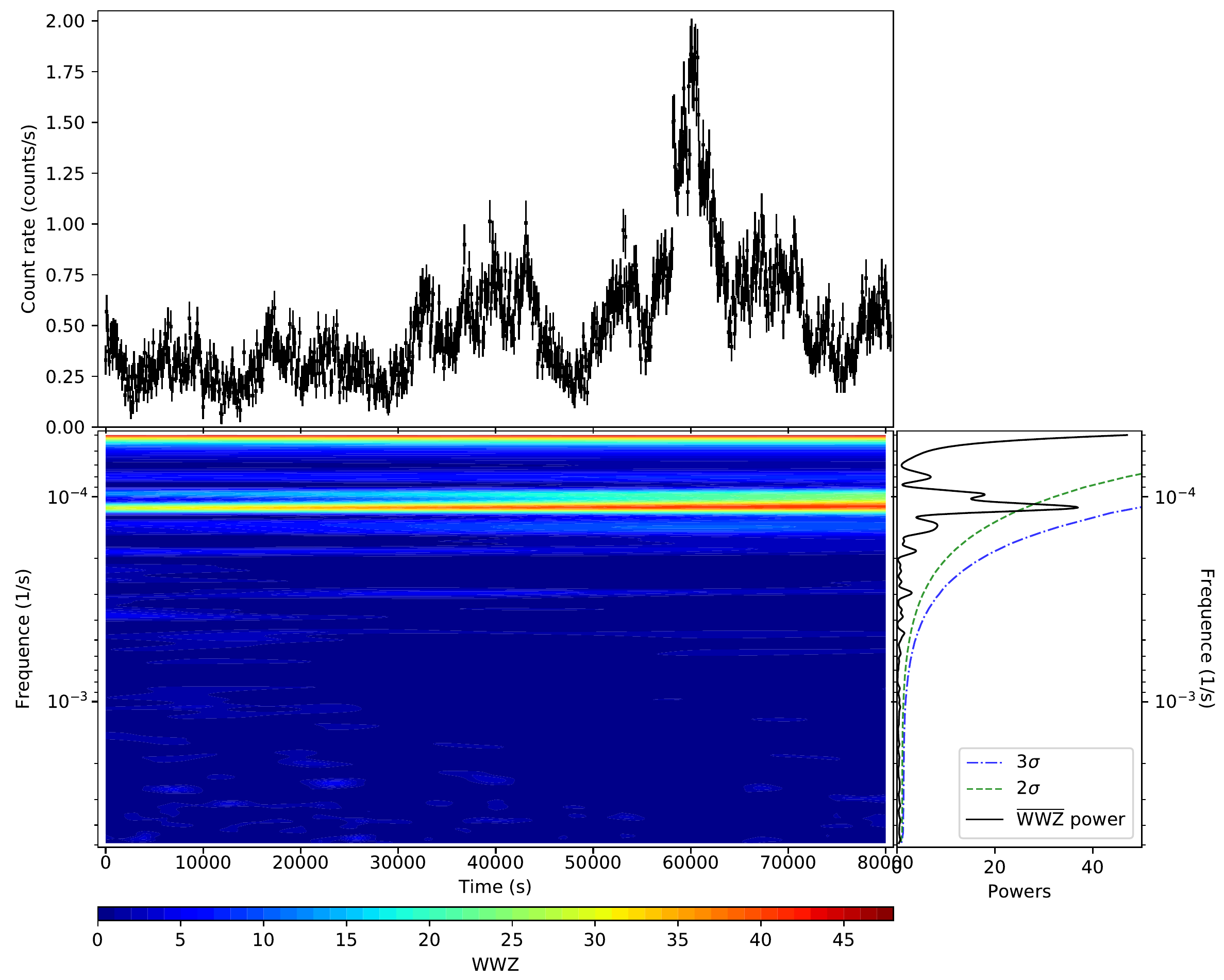}
\caption{The data is for observations of XMM-Newton on 2010 September 15 (Obs ID: 0653510401).
              The confidence curves of $3\sigma$ and $2\sigma$ are shown with blue dotted-dashed and green dashed lines, respectively.
              The parameters $\alpha$ and $Log(f_{bend})$ of the bending power-law model are  $\sim 2.6$ and $\sim -3.9$ for EPIC light-curve and $\sim 2.7$ and $\sim -3.9$ for 2RGS light-curve, respectively.}
\label{20100915}
\end{figure*}

\begin{figure*}
\centering
\includegraphics[scale=0.35]{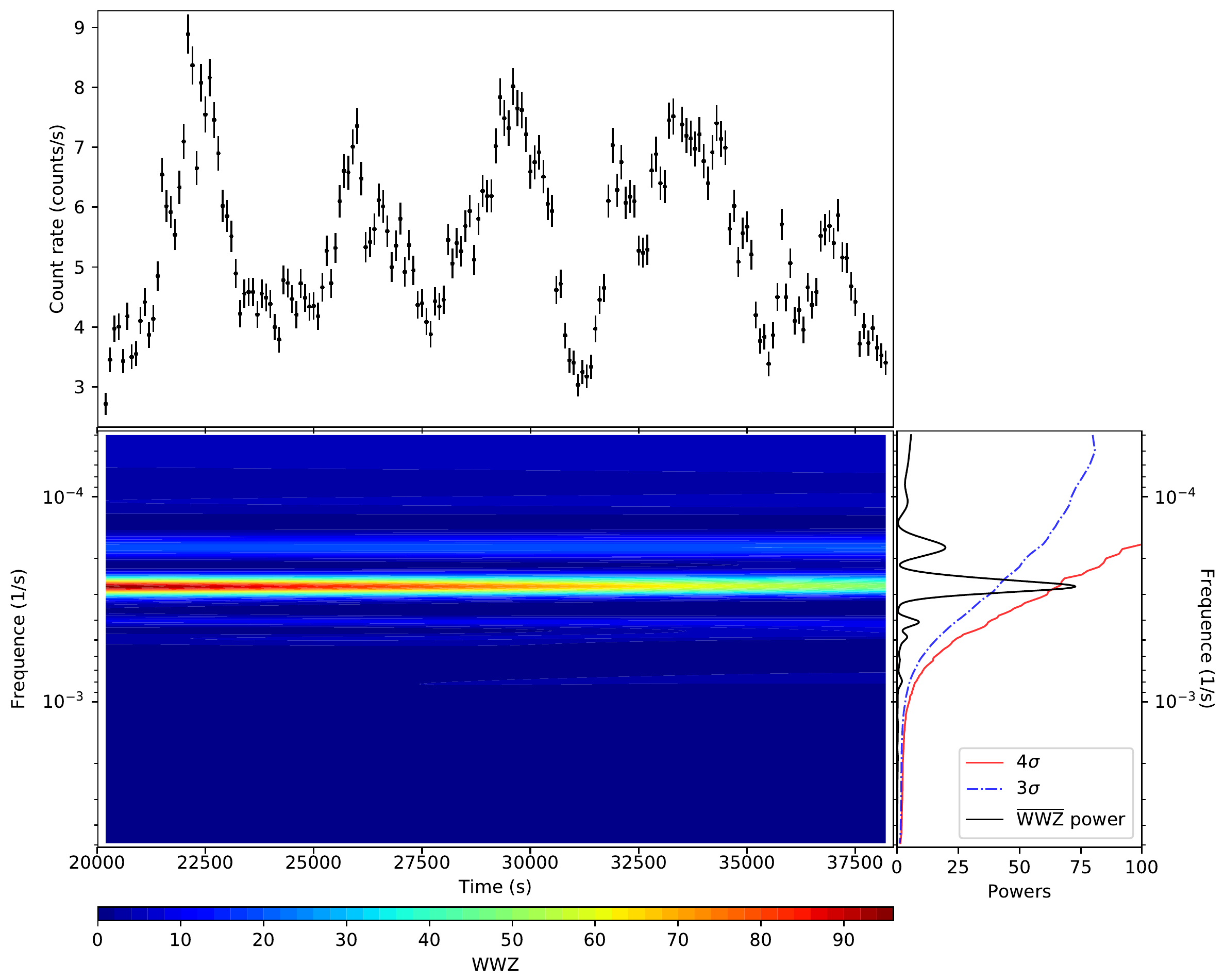}
\includegraphics[scale=0.35]{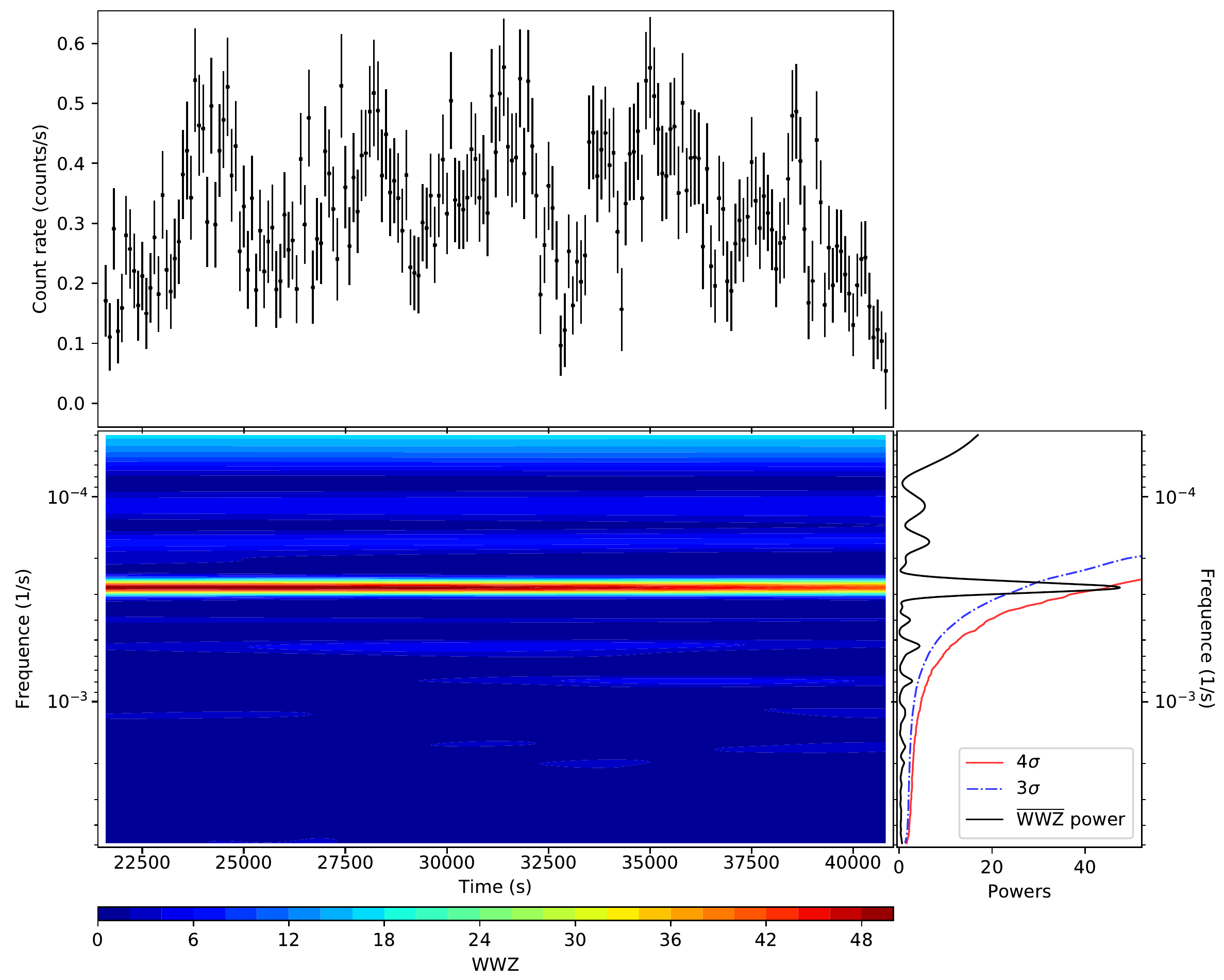}
\caption{The observations of XMM-Newton on 2007 May 14 (Obs ID: 0506200301), left image for  EPIC and right image for 2RGS.
              The confidence curves of $4\sigma$ and $3\sigma$ are shown with red solid and blue dotted-dashed lines, respectively.
              The parameters $\alpha$ and $Log(f_{bend})$ of the bending power-law are  $\sim 2.6$ and $\sim -3.4$ for EPIC
              and $\sim 2.5$ and $\sim -3.7$ for 2RGS.}
\label{20070514}
\end{figure*}

We also fit segment 1 and segment 2 light curves with autoregressive integrated moving average (ARIMA) models \citep{Chatfield2003,Kelly2014,Goyle2017}
to check the reliability of the quasi-periodic signals. Using the Akaike Information Criterion \citep[AIC;][]{Akaike1973},
we select the best-fit ARIMA models for both light curve segments.
And for segment 1 and segment 2, the best-fit models are ARIMA(7,1,7) and ARIMA(9,1,9), respectively.
In the auto-correlation function (ACF) of the residuals of ARIMA(7,1,7) and ARIMA(9,1,9), the most distinct spikes, both exceeding the 95\% confidence limit,
are at a lag of 7400 s for segment 1 and at 3700 s for segment 2, respectively.
The corresponding frequencies are $1.35\times10^{-4}$ Hz and $2.7\times10^{-4}$ Hz, well in agreement with those found with the WWZ method and
indicating that the x-ray quasi-periodic variabilities in segment 1 and segment 2 may be intrinsic.
For segment 1 light curve, the AIC results of 289 ARIMA models are shown in Fig.~\ref{arima_aic} with color-scaled AIC values,
and the standard residuals and auto-correlation function (ACF) of the residuals of the best-fit model ARIMA(7,1,7) are shown in Fig.~\ref{arima}.
Comparing with the LSP and WWZ methods, the ACF can not exactly determine the frequencies of quasi-periodic signal.
Moreover, it is worth noting that the ACF is obtained by the residuals of the best-fit ARIMA model, rather than X-ray light-curve used in methods of LSP and WWZ.
These two reasons may explain the slight difference between the frequency found in ACF and the other two methods.
Furthermore, we know that higher order ARIMA models often produce periodicities in time series.
Then the confidence level of ACF peak obtained by residuals between x-ray light-curves and ARIMA(7,1,7) model will be lower comparing with that of LSP and WWZ.
Even so, the most distinct spike shown in left panel of Fig.~\ref{arima} is well above the threshold of 95\%.

For revealing the variability of X-ray flux, we fold the light-curve with a tool {\it efold} provided in HEASOFT Software using the period cycle of 8244.36 s
with phase zero corresponding to 318550671.023 s. And the tool is provided in HEASOFT Software\footnote{https://heasarc.gsfc.nasa.gov/docs/software/lheasoft/download.html}.
The folded X-ray light-curve is fitted with a constant model, the reduced $\rm\chi^2/d.o.f$ is 7875/49,
it is shown in the I A panel (i.e., the insert) of Fig.\ref{wwz} with a red dashed-dotted line representing the mean count rate of 7.51 counts/s,
which reveals a significant variability of folded light-curve varying with phase.
Furthermore, this result represents the amplitude of X-ray flux varying with phase.
And error bars in the fold light-curve are calculated from the standard deviation of the mean values of each phase bin.
For clarity, we show two period cycles.

We have searched for possible signals in other observation data sets for this source.
Interestingly in two observations there are tentative QPO {signals,
and the results are shown in Figs.~\ref{20100915} and \ref{20070514}.}
For the measurement on 2010 September 15,
the confidence level for the signal in the  EPIC and 2RGS light-curves at $\sim(1.13\pm0.12)\times10^{-4}$ Hz is $\sim 2.4\sigma$  (see Fig.~\ref{20100915}).
While in the left (right) image of Fig. \ref{20070514}, the data is obtained from the measurement on 2007 May 14 for {EPIC} (for 2RGS),
the confidence level with power peak at $\sim(2.70\pm0.24)\times10^{-4}$ Hz is $\sim4.2\sigma$ ($4.1\sigma$).
Furthermore, in the segment 2 of Fig. \ref{wwz_rgs}, we also detect the qusi-peroidic signal at $\sim2.6\times10^{-4}$ Hz in the 2RGS data at confidence level of $\sim 4.6\sigma$,
which is consistent with the results reported in \citet{Pan2016}.
The above periodic cycles are well consistent with those displaying in the emission on
2008 February 4, strongly favoring the transient nature of the QPOs in X-rays of NLS1 galaxies suggested in \cite{Gierlinski2008} and \cite{Pan2016}.

\section{SUMMARY AND DISCUSSION}
\label{sec:summary}

In this work, we have re-analyzed the XMM-Newton observation data of NLS1 galaxy 1H 0707-495.
By dividing the X-ray light-curve measured on 2008 February 4 into two segments, we construct the WWZ powers with the WWZ method, respectively.
In the power spectrum of segment 2 of Fig. \ref{wwz}, there is a strong signal peaks at $(2.64\pm0.2)\times10^{-4}$ Hz,
which confirms the previous detection of QPO at $(2.6\pm0.18)\times10^{-4}$ Hz in \citet{Pan2016}.
Surprisingly, we find a new QPO signal in the power spectrum of segment 1.
Such a signal is at $(1.21\pm0.20)\times10^{-4}$ Hz with a combined significance (from two independent observations) of $\sim 4\sigma$ and
the root-mean-square (rms) in segment 1 has a fractional variability of $\sim30\%$ with a mean count rate of 7.5 counts/s.
On 2008 February 4, the two QPO signals are separated by an intermediate state in light-curve and the frequency ratio is $\sim1:2~(1:2.14\pm0.38)$.
And the signals detected in WWZ power spectra are confirmed with method of ARIMA.
This is the first time to observe two QPO signals, separated by an intermediate state, in X-ray emission of AGNs.
Our conclusion is further supported by the presence of these two signals,
though at lower significance levels, in XMM-Newton measurements on 2007 May 14 and 2010 September 15, respectively.

\begin{figure*}
\centering
\includegraphics[scale=0.6]{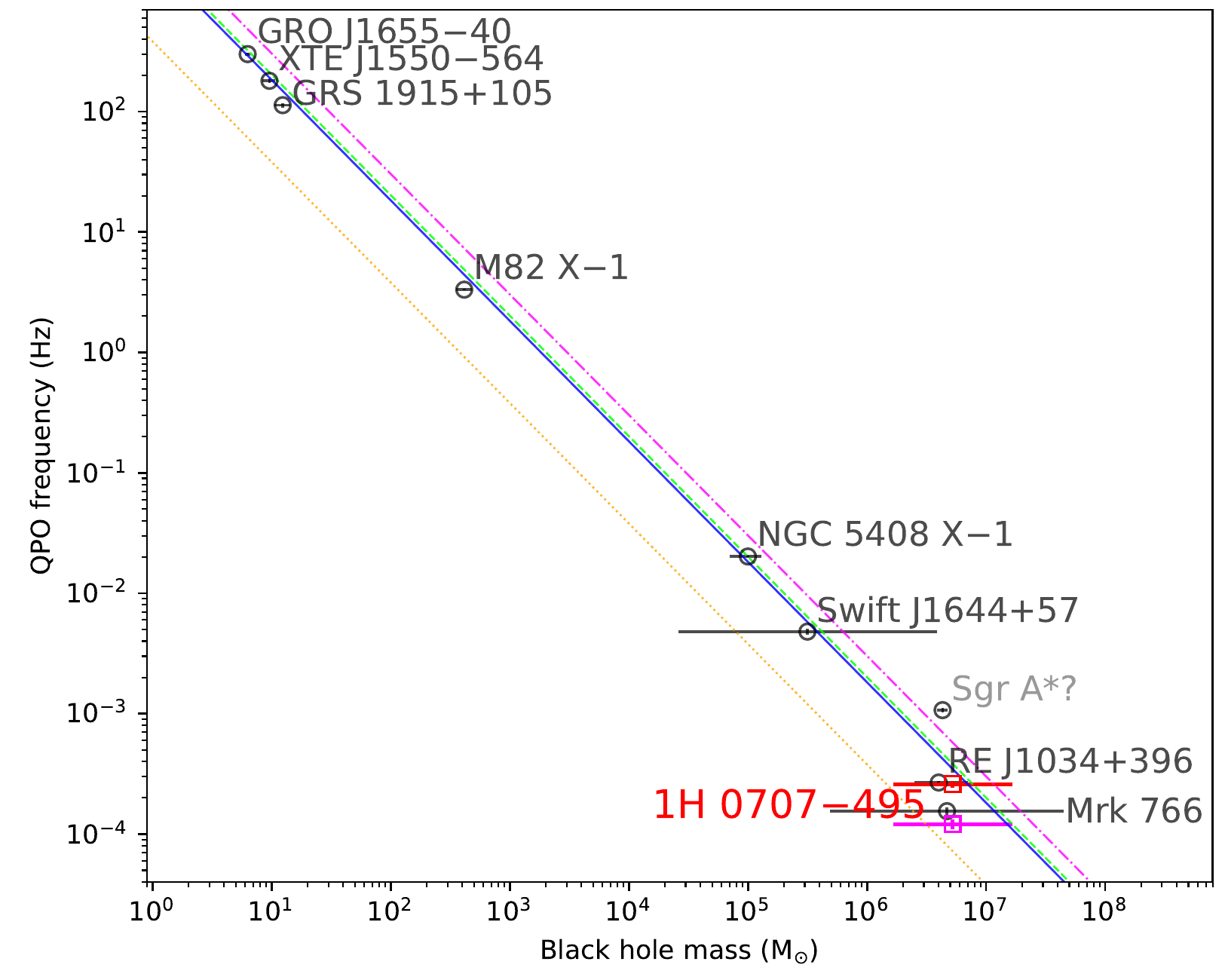}
\caption{The correlation between the BH masses and QPO frequencies. The results reported in previous works are shown with black circles.
              And signals claimed in \citet{Pan2016} and our work are shown with red and purple squares, respectively.
              The detailed information about this relation reference \citet{Kluzniak2002,Abramowicz2004,Torok2005,Remillard2006,Zhou2010,Zhou2015}, please.}
\label{fm}
\end{figure*}

\begin{figure*}
\centering
\includegraphics[scale=0.302]{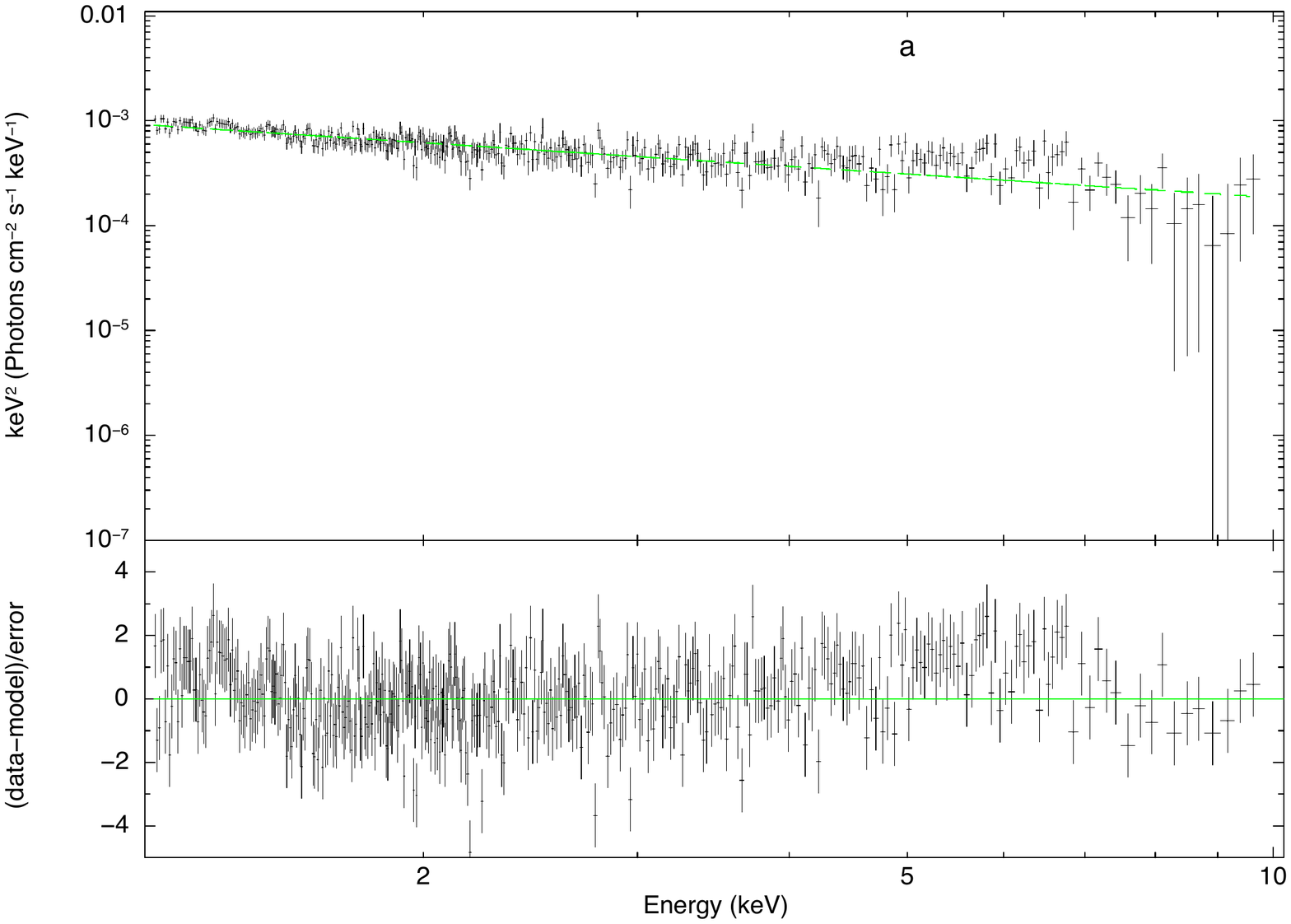}
\includegraphics[scale=0.302]{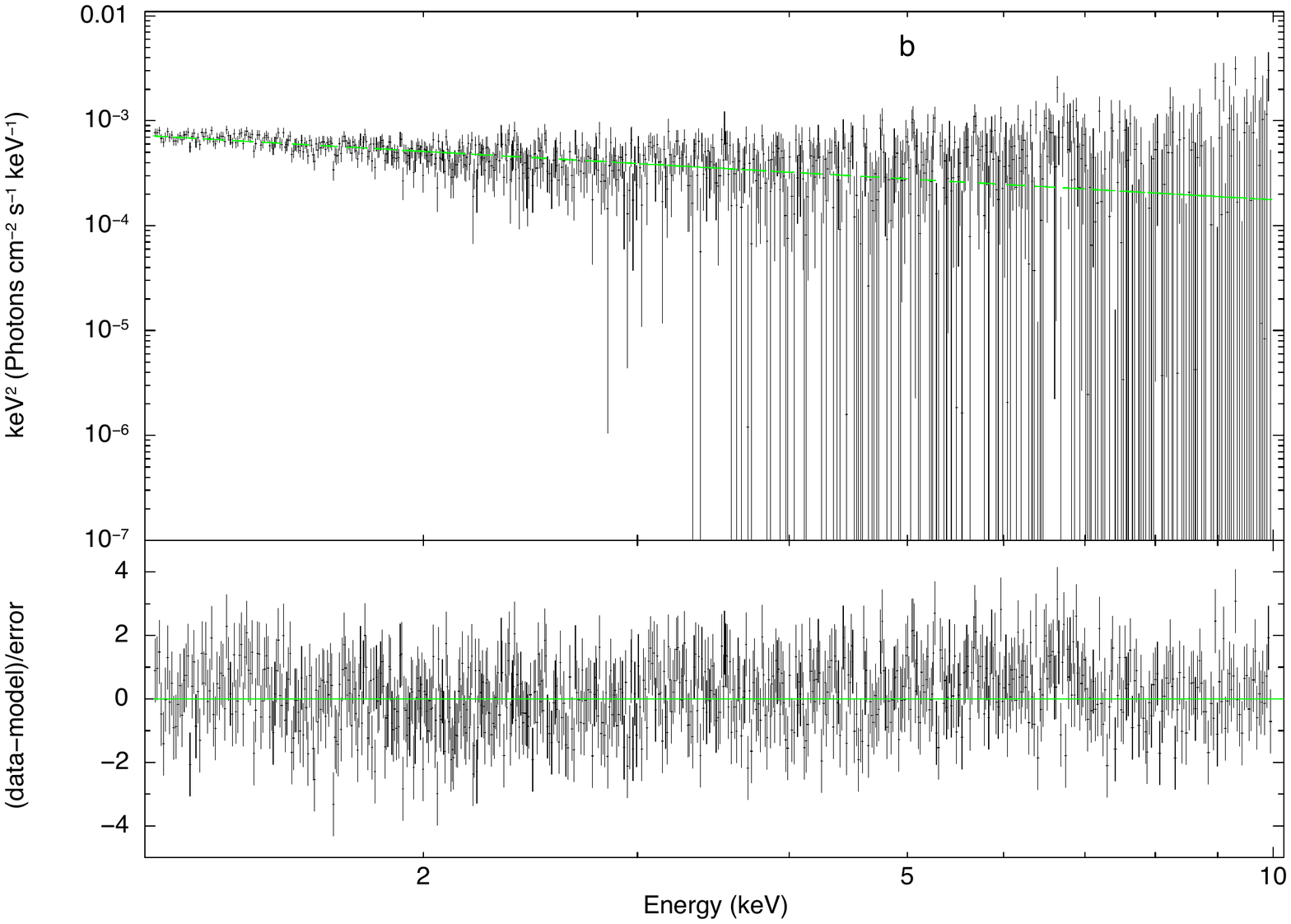}
\includegraphics[scale=0.302]{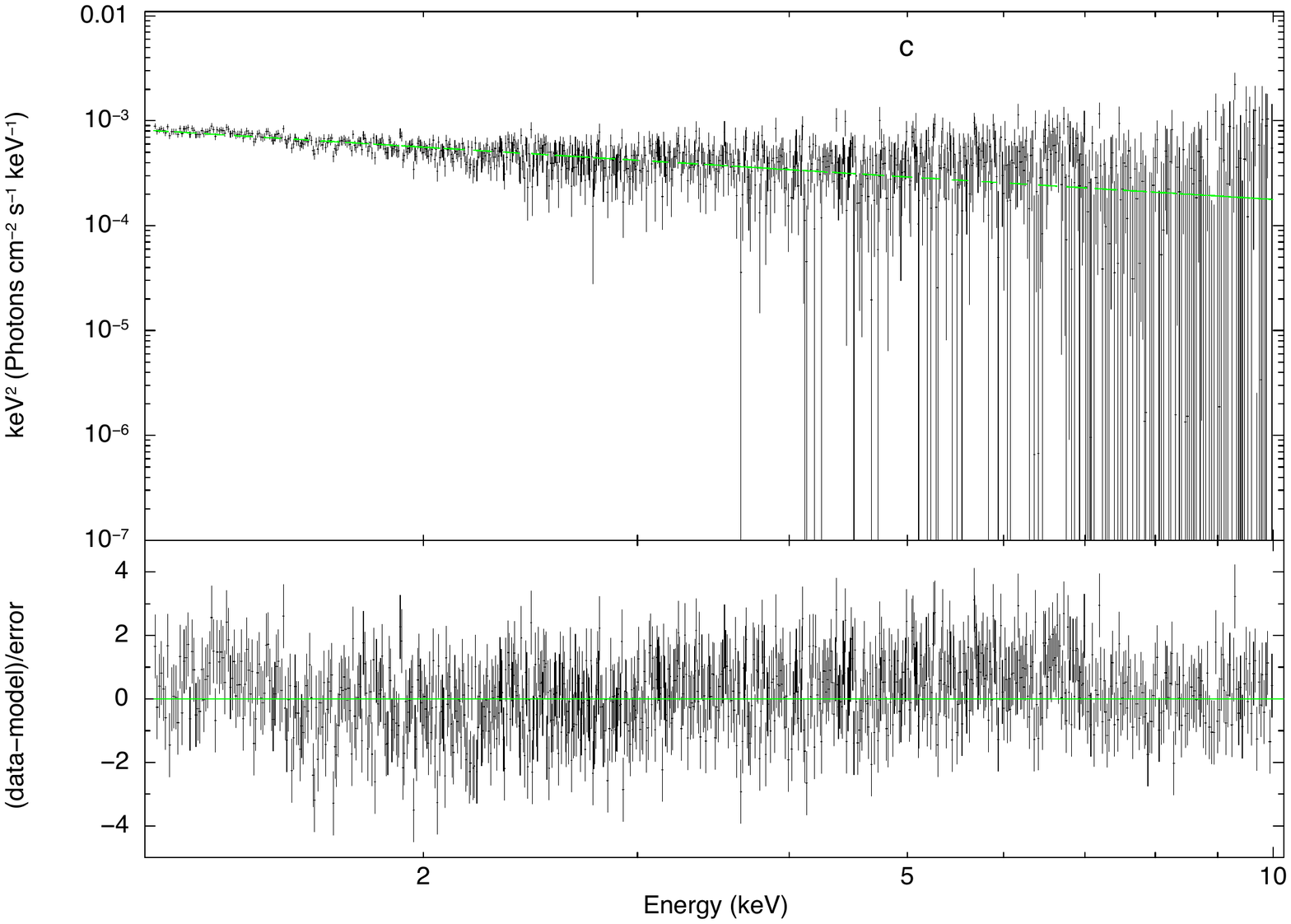}
\caption{The energy spectra from PN cameras on 2008 February 4.
               The fitting results of segment 1 (0 - 44850 sec), segment 2 (44850 - 101800 sec) and time-averaged (0 - 101800 sec) spectra are shown in a, b and c panels, respectively.
               The black crosses and green solid lines represent data and best-fitting model. And energy spectra and residuals are shown in top and bottom for each panel, respectively.}
\label{20080204spec}
\end{figure*}

The physical origin of QPO signals in X-ray binaries as well as AGNs is still to be better understood \citep{Li2004,Remillard2006}. Nevertheless
some models do suggest frequency ratios of $1:2:3$ and so on. For example, \citet{Lai2009} studied the global stability of non-axisymmetric p modes (also called inertial-acoustic
modes) trapped in the innermost regions of accretion discs around black holes and showed that
the lowest-order p modes, with frequencies $\omega \approx 0.6 m\Omega_{\rm ISCO}$ can be overstable due to general relativistic effects, where
$m=1, 2, 3, . . .$ is the azimuthal wavenumber and $\Omega_{\rm ISCO}$ is the disc rotation frequency at the so-called
innermost stable circular orbit (ISCO).
They also suggested that overstable non-axisymmetric p modes driven by the corotational instability may account for the high-frequency
QPOs observed from a number of BH X-ray binaries in the very high state while the absence of such signals in
the soft (thermal) state may result from mode damping due to the radial infall at the ISCO.
While in our scenario it is required that different $m$ appeared in different time intervals.
Our new signal is consistent with the correlation between BH masses and QPO frequencies
\citet{Kluzniak2002,Abramowicz2004,Torok2005,Remillard2006,Zhou2010,Zhou2015,Pan2016}, as shown in Fig.~\ref{fm}.
We extracted the energy spectra of the segment 1, segment 2 and the whole X-ray light-curves on 2008 February 4 using XSPEC \citep[v. 12.9n,][]{Arnaud1996}. The energy spectra are fitted with the model of  \emph{zpowerlaw} (power-law corrected by redshift $\sim0.04$ of 1H 0707-495) and no significant change is found in our
fitting (i.e., they have similar X-ray luminosity and spectral shape). The spectral-fitting results are shown in Fig.~\ref{20080204spec} and the best-fitting parameters are listed in Tab.~\ref{Paras}. Therefore the physical origin of our phenomena is still a mystery.
Finally we would like to caution that the identified QPOs are at relatively low significance and more robust signals are still needed to establish this kind of phenomena.

\section*{Acknowledgments}
We would like to sincerely thank the anonymous referees for the useful and constructive comments.
This work was supported in part by the National Basic Research Program of China (No. 2013CB837000)
and the National Key Program for Research and Development (2016YFA0400200),
the National Natural Science Foundation of China under grants of No. 11525313
(i.e., the Funds for Distinguished Young Scholars), 11433009, 11573071, 11673067, 11733009, U1738124,
the Key Laboratory of Astroparticle Physics of Yunnan Province (No. 2016DG006),
and China postdoctoral science Foundation (No. 2017M621859).

\clearpage
\begin{table*}
 \centering
 \caption{The X-ray spectral-fitting results.}
\begin{center}
\scalebox{0.9}{%
\begin{tabular}{c|c|c|c|c}
\hline\hline
\multirow{2}{*}{} & \multirow{2}{*}{Parameters}& Segment 1 & Segment 2 & Average Spectrum \\ [0.01cm]
 \cline{3-5}
                                                      & & $(0 - 44850)$ s & $(44850 - 101800)$ s &$(0 - 101800)$ s \\  [0.01cm]
\hline
\multirow{2}{*}{zpowerlw} & $\Gamma$                             & $2.74\pm0.02$ &$2.66\pm0.03$ & $2.71\pm0.02$   \\ \cline{2-5}
                                         & $Norm_{\rm zpl}\ (\times10^{-3})$         & $1.15\pm0.02$ &$0.90\pm0.02$& $1.02\pm0.01$   \\
\hline
Flux ($\times10^{-12}$)    & |                                                           & $1.54_{-0.01}^{+0.02}$ & $1.31\pm0.02$& $1.41_{-0.01}^{+0.02}$  \\
\hline
Reduced $\chi^2$ / $\nu$  & |                                                           & 1.29/418           & 1.10/686          & 1.21/822  \\
\hline\hline
\end{tabular}}\\
\end{center}
{{\bf Note:} {The results of X-ray spectral-fitting of segment 1, segment 2 and time-averaged spectra. The redshift in fitting is fixed at 0.04057 and $\Gamma$ is photon index of zpowerlw.
$Norm_{\rm zpl}$ (the normalization of zpowerlw) is in units of photons $\rm cm^{-2}~s^{-1}~keV^{-1}$ at 1 keV and the Flux is in units of $\rm ergs/cm^2/s$.}}
\label{Paras}
\end{table*}

\begin{thebibliography}{}
\bibitem[Abdo et al. (2009)]{Abdo2009} Abdo, A.~A., Ackermann, M., Ajello, M., et al., 2009, ApJ, 699, 976
\bibitem[Abramowicz et al. (2004)]{Abramowicz2004}Abramowicz, M. A., Klu\'zniak, W., McClintock, J. E., \& Remillard, R. A. 2004, ApJL, 609, 63
\bibitem[Ackermann et al. (2012)]{Ackermann2012}Ackermann, M., Ajello, M., Ballet, J., et al., 2012, Science, 335, 189
\bibitem[Ackermann et al. (2015)]{1553}Ackermann, M., Ajello, M., Albert, A., et al., 2015, ApJL, 813, L41
\bibitem[Akaike (1973)]{Akaike1973}Akaike, H. 1973, Biometrika, 60, 255
\bibitem[Arnaud (1996)]{Arnaud1996}Arnaud, K. A., 1996, ASPC, 101, 17
\bibitem[Barret \& Vaughan (2012)]{Barret2012}Barret, D., \& Vaughan, S. 2012, ApJ, 746, 131
\bibitem[Bhatta et al. (2016)]{Bhatta2016}Bhatta, G., Zola, S., Stawarz, \L., et al., 2016, ApJ, 832, 47
\bibitem[Bolton (1972)]{Bolton1972}Bolton, C.T., 1972, Nature, 235, 271
\bibitem[Chatfield (2003)]{Chatfield2003}Chatfield, C. 2003, The Analysis of Time Senes (6th ed.; New York: Chapman \& Hall)
\bibitem[Covino et al. (2017)]{Covino2017}Covino, S., Sandrinelli, A. \& Treves, A., 2017, arXiv:1702.05335v1
\bibitem[den Herder et al. (2001)]{Herder2001}den Herder, J. W., Brinkman, A. C., Kahn, S. M., et al., 2001, A\&A,  365, L7
\bibitem[Emmanoulopoulos et al. (2013)]{Emmanoulopoulos2013}Emmanoulopoulos D., McHardy I. M., Papadakis I. E., 2013, MNRAS, 433, 907
\bibitem[Foster (1996)]{Foster1996}Foster G., 1996, AJ, 112, 1709
\bibitem[Gierli{\'n}ski et al. (2008)]{Gierlinski2008}Gierli{\'n}ski, M., Middleton, M., Ward, M., \& Done, C., 2008, Nature, 455, 369
\bibitem[Goodrich (1989)]{Goodrich1989}Goodrich, R. W., 1989, ApJ, 342, 224
\bibitem[Gonz\'alez-Mart\'in \& Vaughan (2012)]{Gonzales2012} Gonz\'alez-Mart\'in, O., Vaughan, S.,  2012, A\&A, 544, A80
\bibitem[Goyle et al. (2017)]{Goyle2017}Goyle, A., Stawarz, \L, Zola, S. et al., 2017, arXiv:1709.04457
\bibitem[Groth (1975)]{Groth1975}Groth, E. J. 1975, ApJS, 29, 285
\bibitem[Kelly et al. (2014)]{Kelly2014}Kelly, B. C., Becker, A. C., Sobolewska, M., et al., 2014, ApJ, 788, 33
\bibitem[Kidger et al. (1992)]{Kidger1992}Kidger, M., Takalo, L., \& Sillanpaa, A., 1992, A\&A, 264, 32
\bibitem[Kluzniak \& Abramowicz (2002)]{Kluzniak2002}Kluzniak, W., \& Abramowicz, M. A. 2002, arXiv:astro-ph/0203314
\bibitem[Komossa (2008)]{Komossa2008} Komossa, S., Rev. Mex. 2008, Astron. Astrophys. Conf.erence Ser., 32, 86
\bibitem[Lai \& Tsang (2009)]{Lai2009}Lai, D., \& Tsang, D., 2009, MNRAS, 393, 979
\bibitem[Leahy et al. (1983)]{Leahy1983}Leahy, D. A., Darbro, W., Elsner, R. F., et al. 1983, ApJ, 266, 160
\bibitem[Li \& Narayan (2004)]{Li2004}Li, L.-X., \& Narayan, R., 2004, ApJ, 601, 414
\bibitem[Liao et al. (2015)]{Liao2015} Liao, N.-H., Liang, Y.-F., Weng, S.-S., Gu, M.-F., \& Fan, Y.-Z., 2015, arXiv:1510.05584
\bibitem[Markowitz et al. (2003)]{Markowitz2003}Markowitz, A., Edelson, R., Vaughan, S., et al., 2003, ApJ, 593, 96
\bibitem[McHardy et al. (2006)]{McHardy2006}McHardy, I. M., Koerding, E., Knigge, C., et al., 2006, Nature, 444, 730
\bibitem[Osterbrock \& Pogge (1985)]{Osterbrock1985} Osterbrock, D. E. \& Pogge, R. W., 1985, ApJ, 297, 166
\bibitem[Pan et al. (2016)]{Pan2016}Pan, H.-W., Yuan, W.-M., Yao, S., et al., 2016, ApJL, 819, L19
\bibitem[Pogge (2000)]{Pogge2000} Pogge, R.~W., 2000, New Astron. Rev., 44, 381
\bibitem[Reis et al. (2012)]{Reis2012}Reis, R. C., Miller, J. M., Reynolds, M. T., et al., 2012, Science, 337, 949
\bibitem[Remillard \& McClintock (2006)]{Remillard2006}Remillard, R. A., \& McClintock, J. E., 2006, ARAA, 44, 49
\bibitem[Str\"uder et al. (2001)]{Struder2001}Str\"uder, L., Briel, U., Dennerl, K., et al., 2001, A\&A, 365, L18
\bibitem[Timmer \& Koenig (1995)]{Timmer1995}Timmer, J., \& Koenig, M., 1995, A\&A, 300, 707
\bibitem[T\"or\"ok (2005)]{Torok2005}T\"or\"ok, G., 2005, A\&A, 440, 1
\bibitem[Turner et al. (2001)]{Turner2001}Turner, M. J. L., Abbey, A., Arnaud, M., et al., 2001, A\&A, 365, L27
\bibitem[Valtonen et al. (2006)]{Valtonen2006}Valtonen, M. J., Lehto, H. J., Sillanp\"u, A., et al., 2006, ApJ, 646, 36
\bibitem[Vaughan (2005)]{Vaughan2005}Vaughan, S. 2005, A\&A, 431, 391
\bibitem[Webster \& Murdin (1972)]{Webster1972}Webster, B.L., \& Murdin, P., 1972, Nature, 235, 37
\bibitem[Zhang et al. (2017a)]{Zhang2017a}Zhang, P.-F., Yan, D.-H., Liao, N.-H., Wang, J.-C., 2017a, ApJ., 835, 260
\bibitem[Zhang et al. (2017b)]{Zhang2017b}Zhang, P., Zhang, P.-F.,, Yan, J.-Z., Fan, Y.-Z., Liu, Q.-Z. 2017b, ApJ, 849, 9
\bibitem[Zhou et al. (2010)]{Zhou2010}Zhou, X.-L., Zhang, S.-N., Wang, D.-X., \& Zhu, L., 2010, ApJ, 710, 16
\bibitem[Zhou et al. (2015)]{Zhou2015}Zhou, X.-L., Yuan, W., Pan, H.-W., \& Liu, Z., 2015, ApJL, 798, L5
\end{thebibliography}
\end{document}